\documentclass{PoS}
\newcommand\ignore[1]{}
\title{Radial Quantization for Conformal Field Theories on the Lattice}

\ShortTitle{Radial Quantization for Conformal Field Theories on the Lattice}

\author{\speaker{Richard C. Brower}%
         \thanks{RCB acknowledges support under DOE grants DE-FG02-91ER40676,
DE-FC02-06ER41440, and NSF grants OCI-0749317, OCI-0749202.  RCB has benefited
from conversations with Joseph Minaham.  GTF acknowledges partial support by
the NSF under grant NSF PHY-1100905.  HN acknowledges partial support by the
DOE under grant number DE-FG02-01ER41165. HN is grateful for support under the
Weston visiting scientist program at the Weizmann Institute in the Department
of Physics and Astronomy.  HN has benefited from conversations with Adam
Schwimmer and Micha Berkooz.  We also  thank the Galileo Galilei Institute for
Theoretical Physics for the hospitality and INFN for partial support offered
to us (RCB, GTF) during the workshop "New Frontiers in Lattice Gauge
Theories".}\\
        Boston University\\
        E-mail: \email{brower@bu.edu}}

\author{George T. Fleming\\
       Yale University\\
        E-mail: \email{George.fleming@yale.edu}}

\author{Herbert Neuberger\\
       Rutgers University\\
        E-mail: \email{neuberg@physics.rutgers.edu}}

      \abstract{ We consider radial quantization for conformal quantum
        field theory with a lattice regulator. A Euclidean field
        theory on $\mathbb R^D$ is mapped to cylindrical manifold,
        $\mathbb R\times \mathbb S^{D-1}$, whose length is logarithmic
        in scale separation. To test the approach, we apply this to
        the 3D Ising model and compute $\eta$ for the $Z_2$ odd
        primary operator.
 }

\FullConference{The 30th International Symposium on Lattice Field Theory\\
                 June 24 - 29,  2012\\
                 Cairns, Australia}

\begin{document}

\section{Introduction}

Conformal or near conformal behavior in field theory lie at the heart
of many challenging theoretical and phenomenological problems. For
example models that seek to replace the Higgs mechanism of the
Standard Model with a new strong gauge dynamics at the TeV scale often
invoke conformality as an explanation of the flavor hierarchies.
While lattice gauge theory in principle provides a useful approach to
explore such non-perturbative dynamics, conventional lattice methods
for theories that are parametrically  close to conformal theories are
difficult precisely because of the growing separation of the length
scales between the UV and IR. Here we explore a new technique that
replaces the traditional Euclidean lattice in favor of one suited to
{\em Radial Quantization}. Radial quantization has a long history
starting with the observation that the early covariant quantization of
the 2-d conformal string action was given as a radial quantized system
with the Virasoro $L_0$  operator replacing the Hamiltonian. In 1979
Fubini, Hanson and Jackiw~\cite{Fubini:1972mf} suggested radial
quantization of field theory in higher dimensions and later in 1985
Cardy suggested lattice implementations in general
dimensions~\cite{Cardy:1985}.

For an exactly conformal field theory, the  idea is straight forward.   The flat metric  for
any Euclidean field theory on $\mathbb R^D$ can obviously be expressed in radial co-ordinates,
\begin{equation}
 d^2s = dx^\mu dx ^\mu =  r^2_0 e^{ 2t} ( dt^2 + d\Omega^2_{D-1} ) \; , 
\end{equation}
where $t = \log(r/r_0)$, introducing an arbitrary reference scale $r_0$, and
where $d\Omega^2_{D-1}$ is the metric on the $\mathbb S^{D-1}$ sphere of unit
radius.  However in the case of an exactly conformal
field theory, a local Weyl transformation will also  remove
the conformal factor, $\exp[ 2t]$, from the Lagrangian of
the quantum theory. The resultant theory is
mapped from the Euclidean space $\mathbb
R^D$ to a D dimensional cylinder, $\mathbb R\times \mathbb S^{D-1}$.
A simple intuitive illustration of this map begins with the exact
two point function for a primary operator with dimension $\Delta$, 
\begin{equation}
\langle \phi(x_1)  \phi(x_2) \rangle =  \frac{1}{|x_1 - x_2|^{2 \Delta}} \; ,
\end{equation}
and then converts it to radial form,
\begin{equation}
r^\Delta_1 r^\Delta_2 \langle \phi(t_1, \Omega_1)  \phi(t_2, \Omega_2 ) \rangle
 =   \frac{1}{[r_2/r_1 + r_1/r_2 - 2 \cos(\theta_{12})]^{ \Delta}}\rangle \nonumber \\
\simeq  e^{\textstyle - t  \Delta}
\end{equation}
as $t = \log(r_2) - \log(r_1) \rightarrow \infty$. The factors on the left are the Weyl factors absorbed into
the field redefinition of operators for radial quantization. The angular dependence projected
onto  spherical harmonics give rise to integer spaced descendants: $\Delta_l = \Delta + l$.

Our goal is to develop numerical methods to solve  conformal quantum field theories as the infinite
refinement limit of a lattice regularization on $\mathbb R\times \mathbb
S^{D-1}$. If the action is real, one can solve the latter numerically
by Monte Carlo methods, extract quantitative features and test the
mathematical question of convergence to a universal continuum limit.
If this is possible, a potential advantage is that a lattice with
$T$ sites in $t = \log(r/r_0) $ represents an exponential scale
separation as function of $T$ relative to conventional Euclidean
finite lattice on $\mathbb R^D$.

\section{Lattice Implementation}

\begin{figure}[ht]
\centering
\hfil
\includegraphics[width=0.45\textwidth]{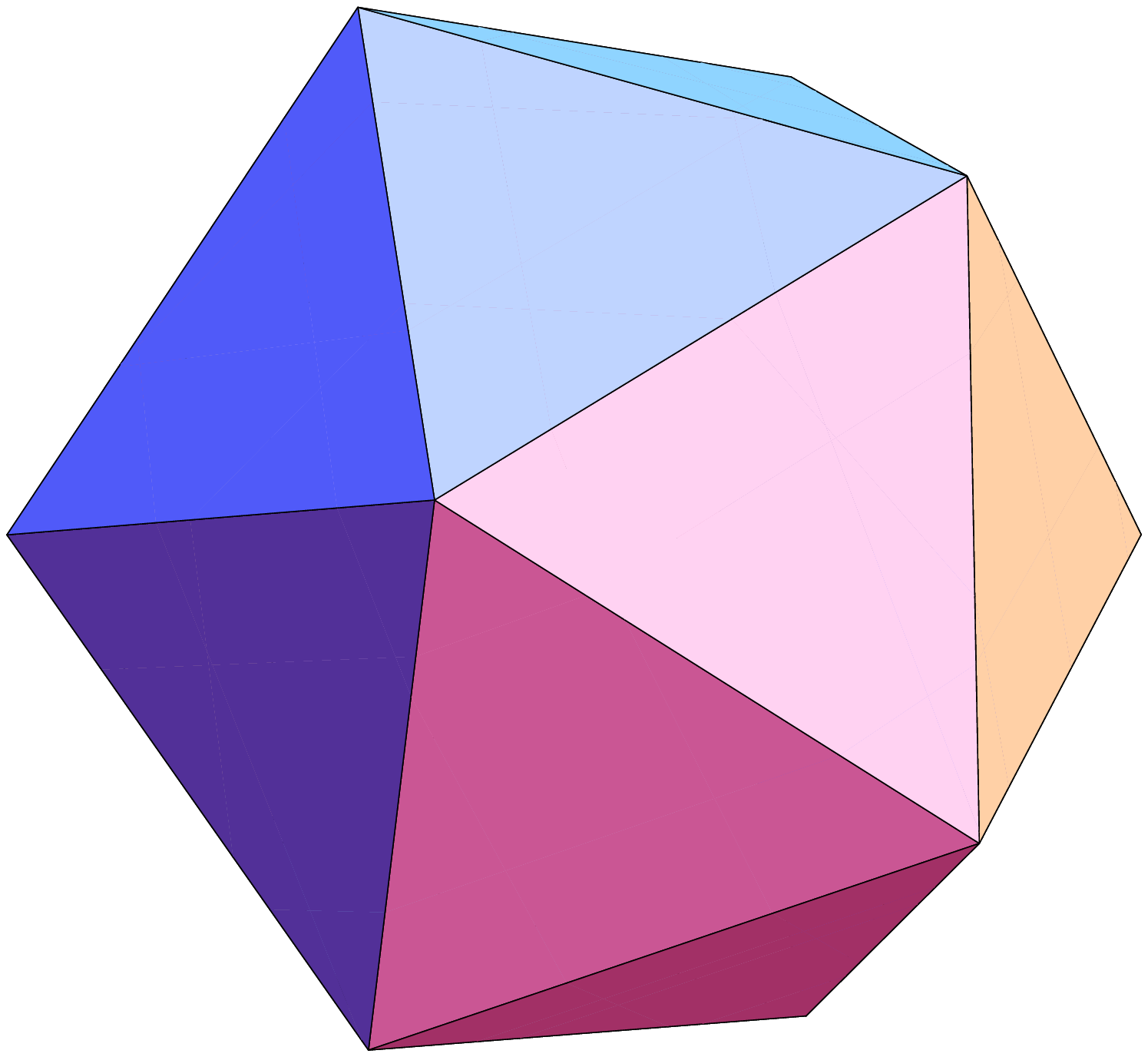}
\hfil
\includegraphics[width=0.45\textwidth]{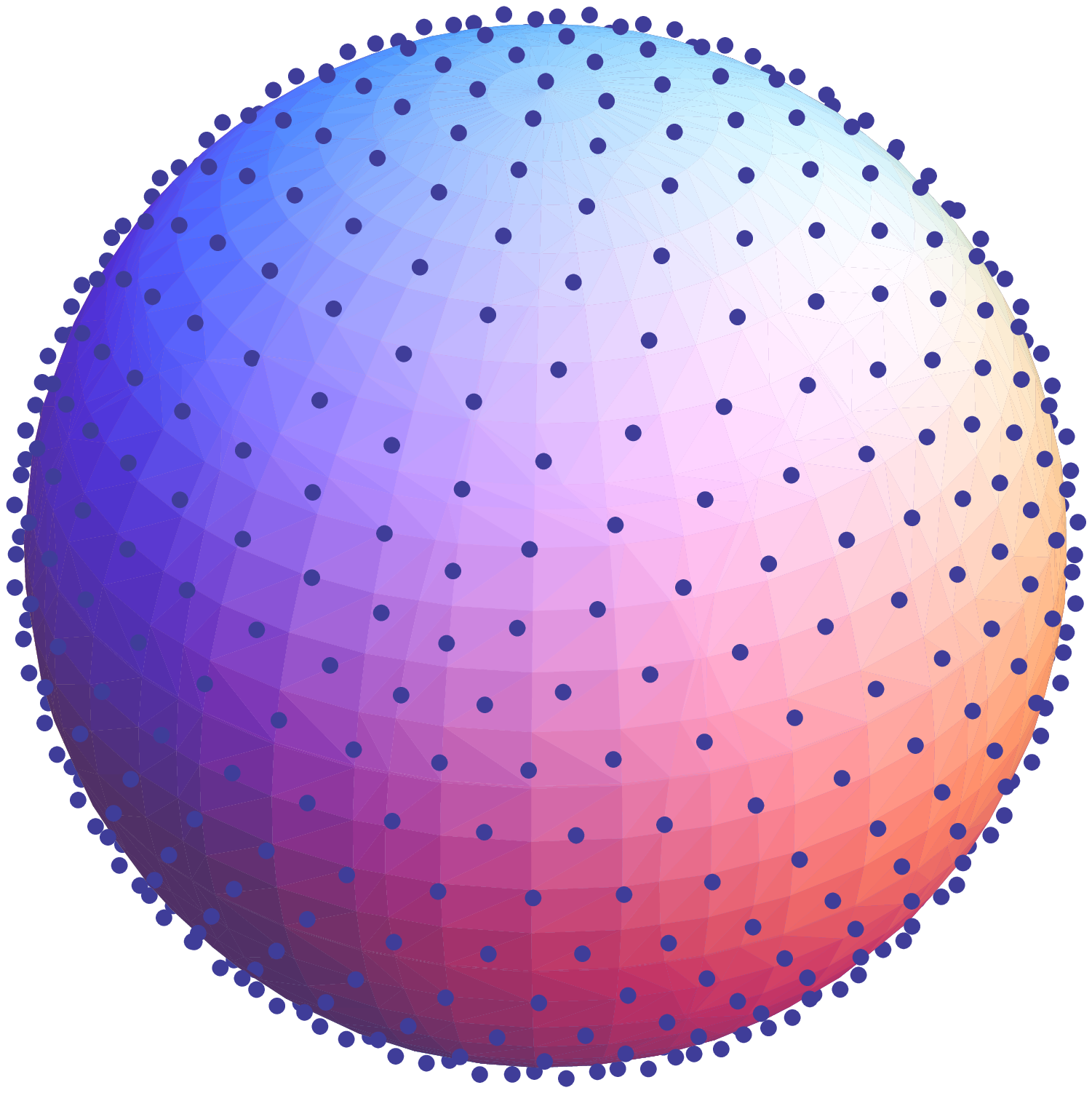}
\hfil
\vskip -1.2 cm
\caption{\label{fig:Icosahedron} On the left is the $s =1$ icosahedral approximation 
to the sphere and on the right the $s = 8$ equilateral triangle refinement of the icosahedron , illustrated by projecting each vertex at fixed angles from the center of the icosahedron to the unit sphere. }
\end{figure}

As a test of this idea, we present first results for the 3D Ising
model at the Wilson-Fisher critical point.    The largest
discrete subgroup of the isometries of $\mathbb R\times \mathbb
S^{2}$ are achieved by a uniform lattice for the non-compact  $\mathbb R$ co-ordinate
  and an icosahedral approximation to the compact sphere, $\mathbb
S^{2}$.  The icosahedron has 12 vertices and 20 faces given by identical flat 
equilateral triangles as illustrated in Fig.~\ref{fig:Icosahedron}.  Its symmetry group $I_h$ is a 120 element subgroup of
$O(3)$. The angular momenta $l=0,1,2$ representations of $O(3)$ remain
irreducible representations under $I_h$.  There is enough symmetry to isolate
a scalar primary state and at least two of its immediate descendant states.
To refine the lattice on the icosahedron 
each face is subdivided into $s^2$ equilateral triangles.  The sites on each icosahedral surface
at fixed $t$ are connected to the corresponding sites 
on the neighboring surfaces at $t \pm 1$.   

The partition function is of the usual form, 
\begin{equation}
Z = Tr_\sigma  e^{\textstyle  \sum_{t,x}  \beta \sigma(t,x) \sigma(t+1,x)+\sum_{t,\langle x y\rangle} \beta  \sigma(t,x)\sigma(t,y)}  \; ,
\end{equation}
where $\langle x y\rangle$ denotes a nearest-neighbor pairs on each
icosahedral shell and $t = 0, \cdots T-1$ sums over the radial co-ordinate. 
The trace is the sum over the Ising spin,
$\sigma(t,x) = \pm 1$, on each site $(t,x)$. For finite $s$, the
logarithm of the transfer matrix along the cylinder is a regularized
representation of the dilatation operator. To get information on the
spectrum of the transfer matrix, it is convenient to compactify the
infinite axis of the cylinder to a circle with periodic boundary
conditions on the spins.

To approach the Wilson-Fisher conformal field
theory in the continuum limit, we need to tune $\beta$ to the  critical point.
The relative scale between the longitudinal and transverse lattice
(``speed of light'' ) is fixed by the integer spacing of descendants
of the primary operators. There are no other free parameters. For
example the leading primary operator odd under $Z_2$ has a sequence of
descendants $\Delta_l = 1/2 + \eta/2 + l$ for $l = 0,1, ... $. The
first 3 states were clearly identified with modest calculations,
verifying the integer spacing and determining the anomalous
contribution $\eta$.

\section{Numerical Results}

The number of sites on one icosahedral shell is $ 2 + 10 s^2 $.
We chose our cylinders to have lengths which scale with the refinement
$T \equiv \rho  s$.  To locate the critical point, we used aspect ratios
$\rho = 4, 8$ while for computation of the magnetization correlation functions
we used only $\rho = 8$.

\begin{figure}[ht]
\centering
\hfil
\includegraphics[width=0.45\textwidth]{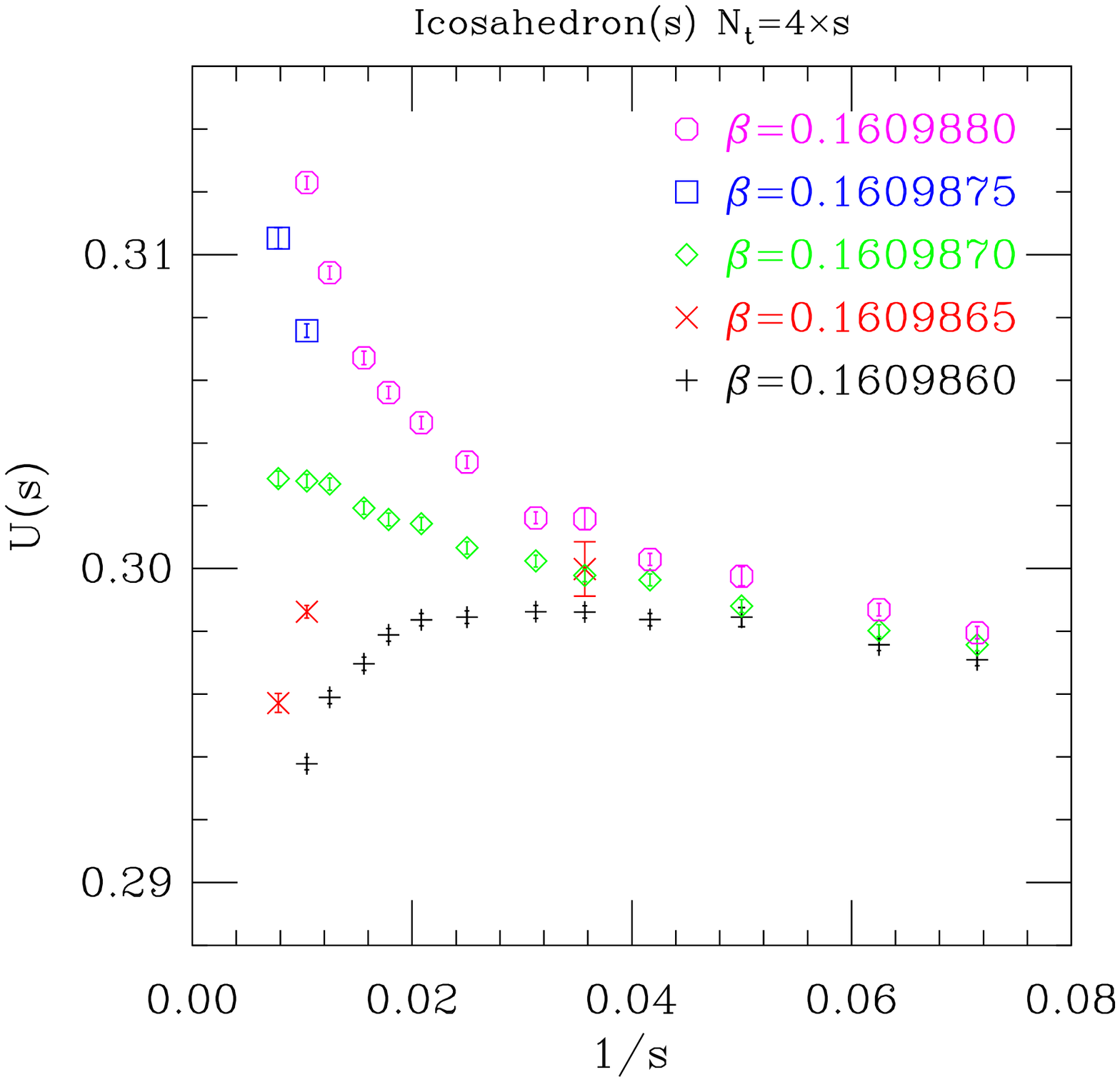}
\hfil
\includegraphics[width=0.45\textwidth]{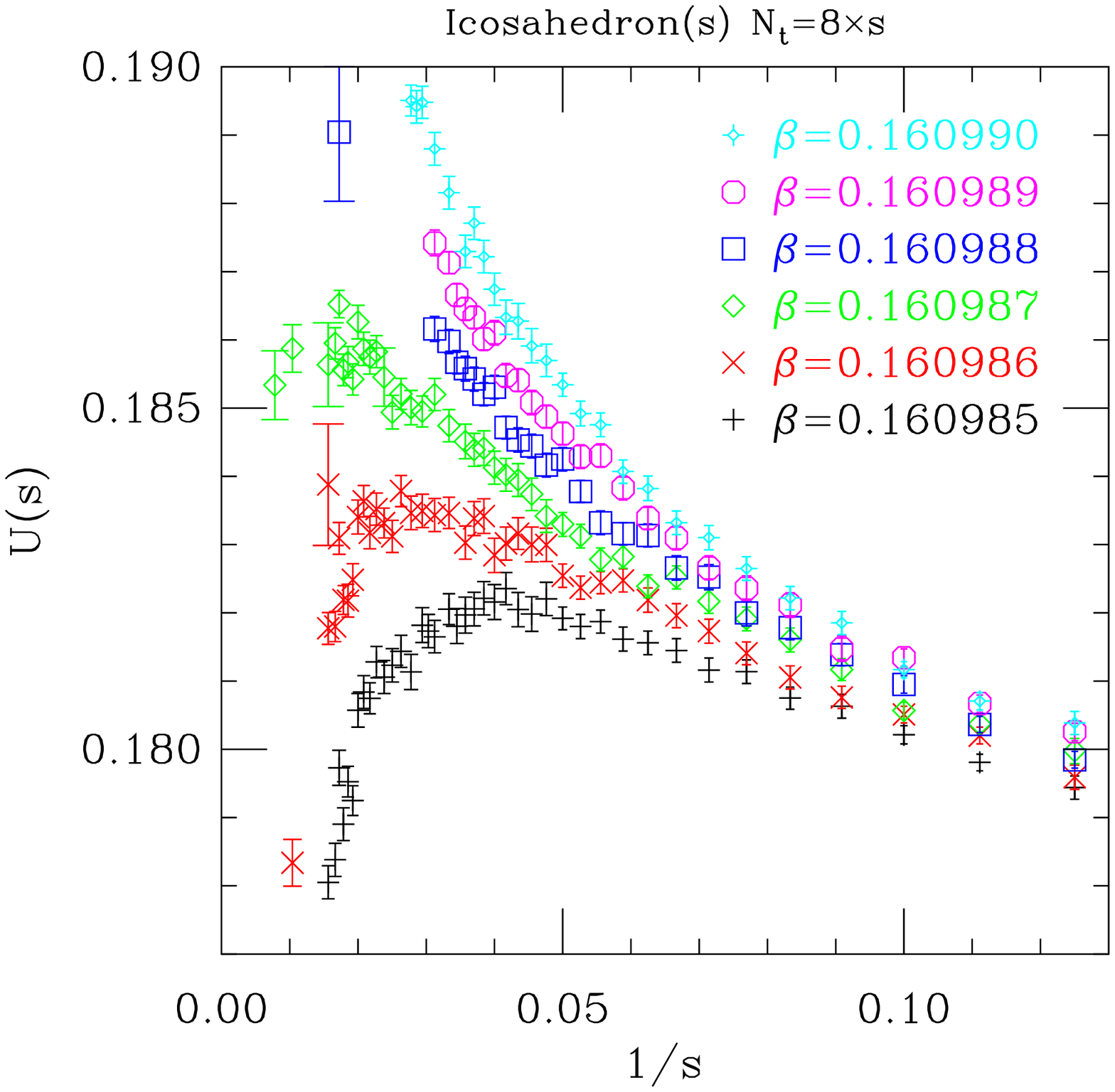}
\hfil
\caption{\label{fig:Binder}Determining $\beta_c$ from the Binder cumulants
$U(s) = 1 - \langle M^4\rangle /3 \langle M^2 \rangle^2$ near the
pseudo-critical point for two different aspect ratios $\rho$ and increasing
values of s.}
\end{figure}

The critical point $\beta_c$ was determined first by constructing a sequence
of pseudo-critical $\beta$-values defined as matching points of the Binder
cumulants\cite{Binder:1981sa},
\begin{equation}
\label{eq:Binder_cumulant}
U(\beta,s,\rho) = 1 - \frac{\langle M^4 \rangle}{3 \langle M^2 \rangle^2}
\end{equation}
for $s_{p+1}=r s_p$ at consecutive $p$-values.  This was done for several
values of $r$ in the range $1.5\le r \le 5$.  We also numerically obtained the
subleading terms in the approach to the fixed point predicted by the
Renormalization Group. Subsequently we improved the estimate by performing a
global fit of the scaling relation,
\begin{equation}
U(\beta,s,\rho) \simeq U(\beta_c,\infty, \rho) + a_1(\rho) [\beta - \beta_c]
s^{1/\nu} + b_1(\rho) s^{-\omega} 
\end{equation}
to many independent simulations ($\rho = 4, 8)$, a subset of which appear in
Fig.~\ref{fig:Binder}, of the scaling relation where the exponents $\nu$ and
$\omega$ were held constrained to agree with the published values
\cite{Pelissetto:2000ek}.  We find $\beta_c$ = 0.16098698(2),
$U(\beta_c,\infty,4)$ = 0.3040(2) and $U(\beta_c,\infty,8)$ = 0.1876(2).

We started our study employing the Swendsen-Wang cluster
algorithm~\cite{Swendsen:1987ce} but switched to the more efficient single
cluster Wolff algorithm~\cite{Wolff:1988uh}.  For our final results on the
spin-spin correlation function, we generated ensembles at $\beta = 0.160987$.
Each independent run was thermalized using 2048 sweeps of the Wolff algorithm
followed by 8192 sweeps with one estimate of the spin-spin correlation
function after each sweep.  We defined a sweep to be $19 s / 2$ Wolff cluster
updates which leads to the average number of spins flipped each sweep equal to
the total volume.  All results for a given run are then averaged together to
form a single blocked estimate and many thousands of independent block
estimates are combined to form the ensemble at each $s$.
% as shown in Table~\ref{tab:runs}.  
The jackknife method was used to estimate errors.

 We projected the spin-spin correlations function on spherical harmonics,
\begin{equation}
C_{l}(t)  = \sum_{m, t_0, x,y}   Y_{lm}(\Omega_x) \; \langle \sigma(t+t_0,x) \sigma(t_0, y) \rangle \;  Y_{lm}(\Omega_y) \end{equation}
where $Y_{lm}(\Omega_x) $  is spherical harmonic evaluated at the
angular position of the site $x$, weighted by 1/3 of the area of the adjacent spherical triangle
projected on the unit sphere as illustrated in Fig.~\ref{fig:Icosahedron}. This represents a finite element approach giving an improved approximation to 
orthonormality of the discrete spherical harmonic.  As we are only interested in the
rotationally-invariant part of the correlation function on any given lattice,
we have summed over the azimuthal quantum number, $m$.

In addition we found it very useful to evaluate  the spin-spin correlation
function using the momentum space single cluster improved estimator method
\cite{Ruge:1994jc}.  The connected correlator for the lowest mass discrete eigenstate of
transfer matrix on our periodic lattice is exactly represented by a single
hyperbolic cosine,
\begin{equation}
C_l(t) = A_l \cosh(- \mu_l(t - T/2))  \; ,
\end{equation}
at discrete values $t = 0, \cdots, T-1$. We transform this to momentum space, 
\begin{equation}
\widetilde C_l(k) = \frac{1}{T} \sum_{t=0}^{T-1} e^{ \textstyle i t k}  \;
C_l(t) = c_0 \delta_{l,0} \delta_{k,0} + a_{l} \frac{(1 - e^{-\mu_{l} T})
\sinh(\mu_{l})}{\sinh^2(\mu_{l}/2) + \sin^2(k/2) } .
\end{equation}
where  $k = 2 \pi q / T$ with $q = 0, \cdots, T-1$ is the momentum conjugate
to $t \equiv \log r $ along the cylinder.  Since our value of $\beta \simeq
\beta_c$ is slightly larger that the pseudo-critical coupling at any finite
$s$, we expect that our $l = 0$ correlation function will have a small
disconnected contribution. This contributes a non-analytic term,
$c_0\delta_{k,0}$. In momentum space the disconnected piece can in principle
be isolated by subtracting a
smooth extrapolation of $C_0(k)$ from $k\ne0$ to $k = 0$.
\begin{figure}[ht]
\centering
\hfil
\includegraphics[width=0.45\textwidth]{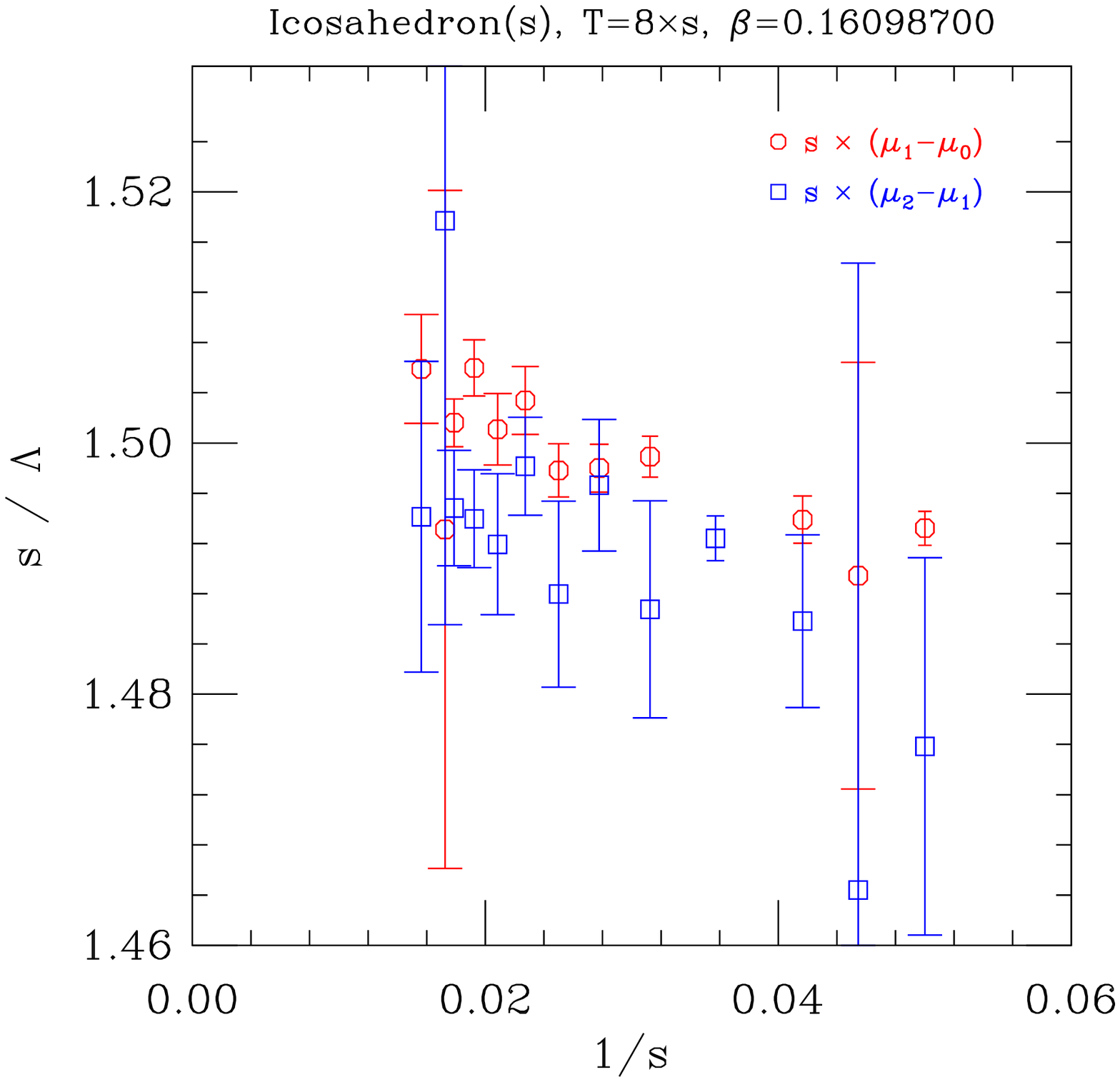}
\hfil
\includegraphics[width=0.45\textwidth]{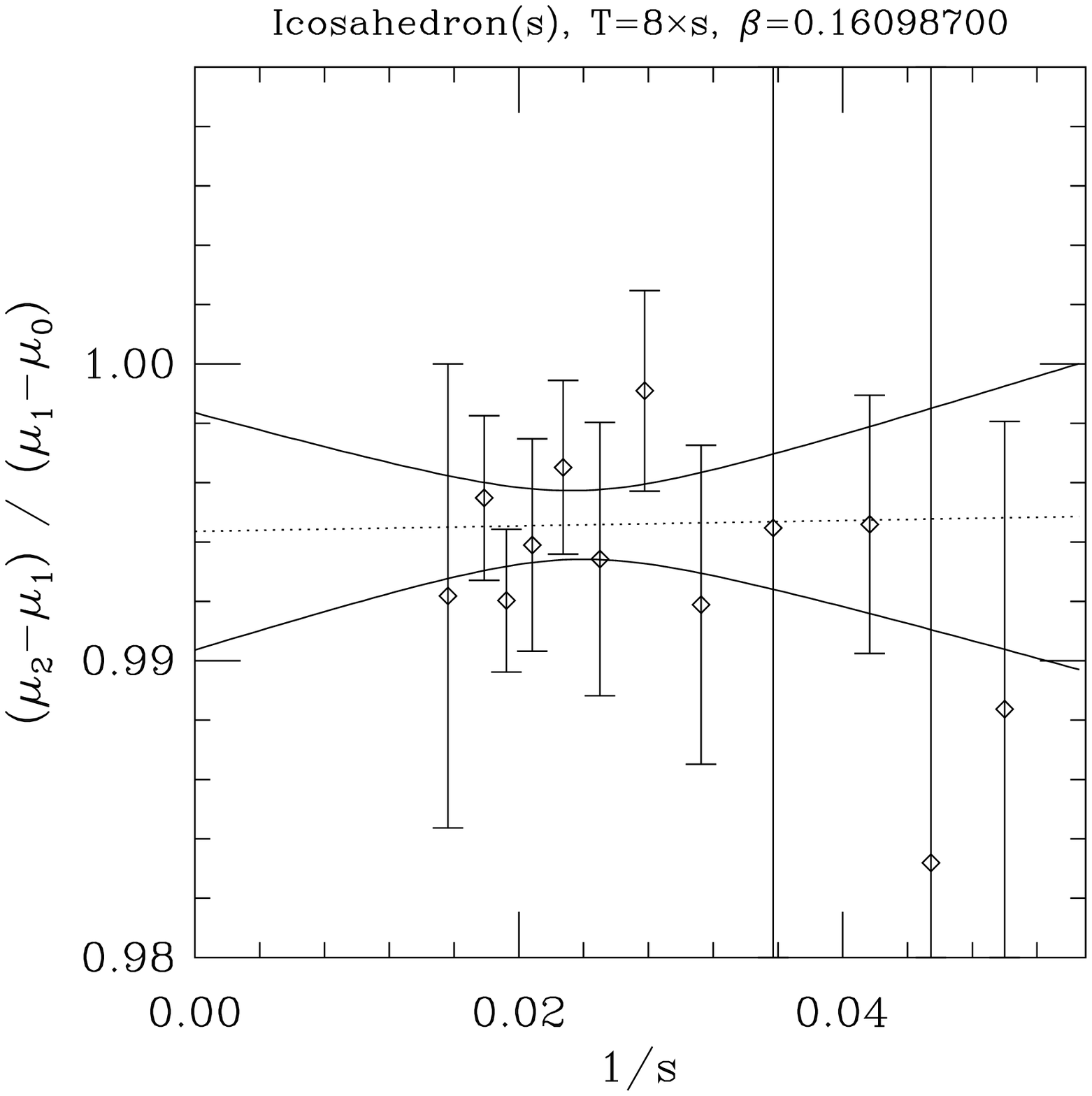}
\hfil
\caption{\label{fig:scaling_spacing} The left figure shows the scaling of
$\Lambda$ relative to $s$.  The extrapolated value is roughly 1.51(1) with the
uncertainty dominated by the systematic difference between the two estimates.
The right figure tests the hypothesis of integer level spacing between
primaries and descendents by comparing the level spacing between the first and
second descendents to the spacing between the primary and its first
descendent.  We fit to a linear function and find an intercept of 0.994(4) and
slope of 0.0(2) with $\chi^2/\mbox{dof} = 0.43$ for 11 dof.}
\end{figure}

We found that our data require parameterizing the ground state plus at least
three higher mass states to get excellent fits with $\chi^2 / \mbox{dof}
\lesssim 1$ and estimates of ground state masses which are essentially free of
higher state contamination.  These represent possible high states either
propagating forward in the $Z_2$ even or propagation backward in $Z_2$ odd
sectors. While the dimension of the higher state coming from the $Z_2$ even
sector might be lower, its mixing will also be proportional to the 3 point
coupling of the energy operator and two spin operators. Indeed with higher
statistics, we believe quantitative determination of the higher spectrum is
well within reach of this method. 
\begin{figure}[ht]
\centering
\includegraphics[width=0.45\textwidth]{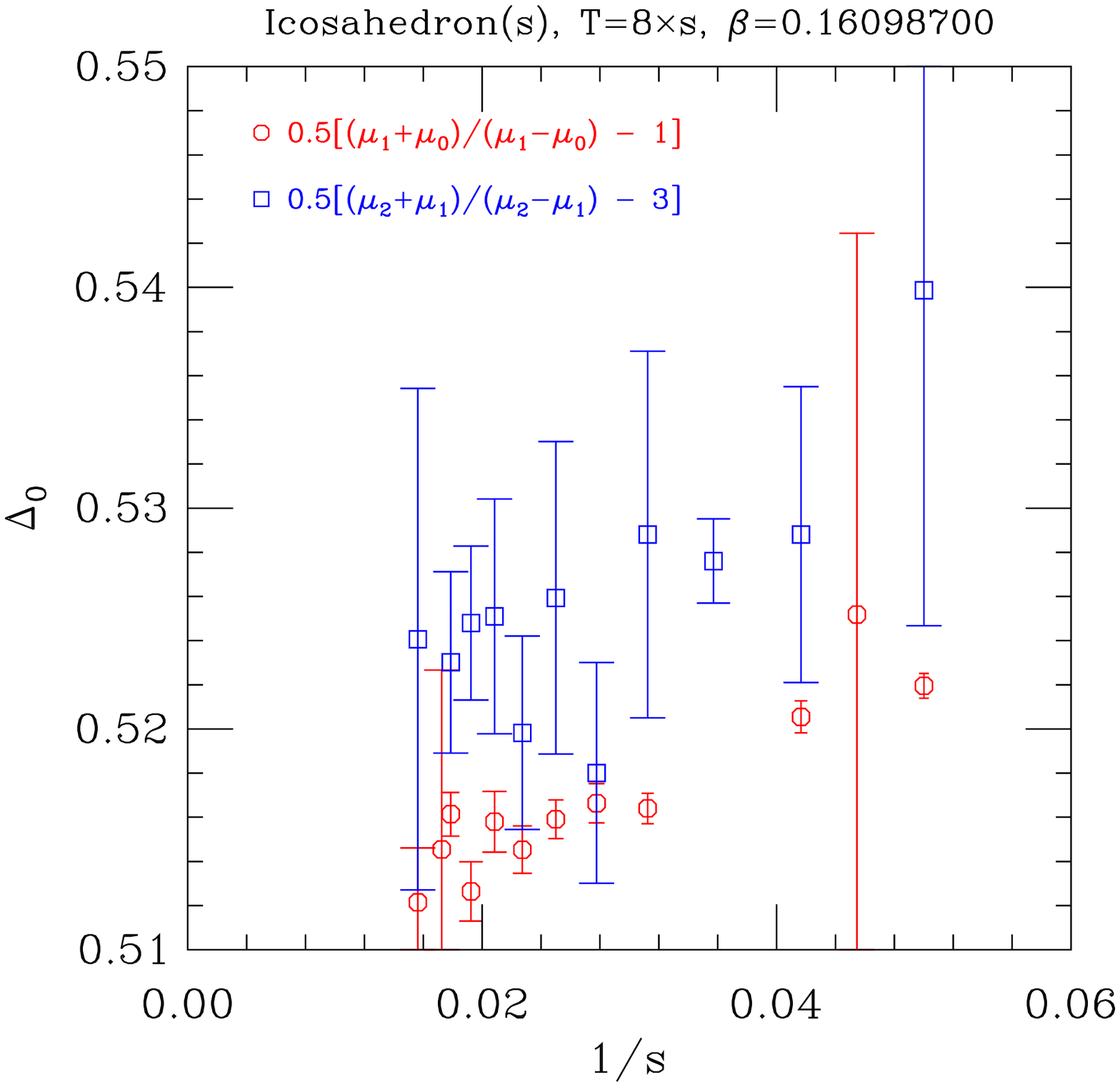}
\caption{\label{fig:delta} The scaling exponent of lowest $Z_2$-odd primary
operator \textit{vs.}\ $1/s$.  The lower (red) points using the primary and first descendant
have disconnected contributions  that have not been fully determined. 
A linear extrapolation from upper (blue) points, using the first two descendants which have no  disconnected contributions,
 gives $\Delta_0 = 0.517(5)$,  is consistent with the best published estimate 0.5182(3) \cite{Pelissetto:2000ek}.}
\end{figure}

Once we have determined the $\mu_l$'s, we relate them to the eigenvalues of
the dilatation operator up to a single  unknown constant: $\mu_l =
\Lambda^{-1} [ \Delta_0 + l ]$ where $\Lambda^{-1} \simeq c_1 / s $ as $s \to
\infty$.  Numerically, we find $c_1 \approx 1.51(1)$ with the uncertainty
dominated by systematic error.  Clearly, we see evidence for sub-leading
contributions $\mathcal{O}(1/s^2)$ as well in the left figure of
Fig.~\ref{fig:scaling_spacing}.  We then test for the equal spacing rule of
descendants  by examining the ratios, $(\mu_{l+2} - \mu_{l+1}) / (\mu_{l+1} -
\mu_l)$, as shown on the right in Fig.~\ref{fig:scaling_spacing}.  Using this
confirmation, we are able to estimate numerically the scaling dimension of the
primary operator using ratios,
\begin{equation}
\Delta_0 = \frac{l-l^\prime}{2} \left[
  \frac{\mu_{l} + \mu_{l^\prime}}{\mu_{l} - \mu_{l^\prime}}
  - \frac{l + l^\prime}{l - l^\prime}
\right]
\end{equation}
as shown in Fig.~\ref{fig:delta}.

We end up with what we feel is a quite conservative estimate of
$\eta=0.034(10)$, consistent with other estimates \cite{Pelissetto:2000ek}.
Moreover this method can be extended to include additional primary operators
in both the $Z_2$-odd and $Z_2$-even sectors as well as direct test of the
restoration of full conformal symmetry for 2- and 3-point correlators.

\section{Discussion}

We have presented a simple example of lattice radial quantization for
the 3D Ising model. This raises important questions and suggests
further applications. 

For the 3D Ising radial quantization, we have chosen a simple
approximation to the sphere by a triangular refinement of the
icosahedron with ``flat'' sides. This is reflected in the action by
assigning equal weights on all the nearest neighbor links. Viewed in
the language of Regge calculus, this geometry has all the curvature
concentrated at 12 exceptional vertices of the underlying icosahedron
that are bounded by 5 rather than 6 triangles. We are implicitly
making the conjecture that by virtue of maintaining exact icosahedral
symmetry the continuum limit gives back the symmetry of $\mathbb S^2$,
indeed the full conformal group. Our modest numerical results to date
support this conjecture, but we are undertaking more stringent
numerical test on the spectrum and correlators. It is an open question
whether the conical singularities at the vertices of the icosahedron
are irrelevant to the continuum spectrum. If necessary 
one might introduce improved triangulation of the metric on  $\mathbb S^2$,
similar to our improvement of the $Y_{lm}$ weights in
  our correlation measurement.  On the other hand, it is also interesting to ask if the
  simpler geometry of a refined cube is adequate, since applications to
  4D gauge theories are probably easier to formulate on concentric 3D
  hypercubes.
 
The next simplest model beyond the 3D Ising model to consider is  the 3D O(N) model, which
because of the analytical results in the large N limit is an excellent
test-bed for the method. We are considering generalization to
include gauge fields and fermions in 3D and 4D.  Each of these steps will
require careful consideration to make sure that there are no
obstructions to taking the continuum limit. There maybe subtle issues
for example with fermions on a spherical manifolds and potentially
relevant conformal symmetry breaking operators the don't vanish in
the continuum.  Even more challenging  are theories that
are not quite conformal where the dilatation operator is no longer a
conserved quantity,  such as those exhibiting asymptotic freedom in the UV 
and softly broken conformality near an IR fixed point.

\end{document}